\documentclass[a4paper,UKenglish]{lipics}
 
\usepackage{microtype}

\newcommand{\aRel}{{\ensuremath{\mathrel{{\mathcal R }}}}}
\newcommand{\ar}[1]{\mathrel{\stackrel{#1}{\longrightarrow}}}
\newcommand{\sset}[1]{\{ {#1}  \}  } 
\newcommand{\Defs}{\mathrel{:=}}
\newcommand{\dar}[1]{\mathrel{\stackrel{#1}{\Longrightarrow}}}

\def \Rm#1{\mbox{\rm #1}}
\def \lsem      {\raise1pt\hbox{\Rm {[\kern-.12em[}}}
\def \rsem      {\raise1pt\hbox{\Rm {]\kern-.12em]}}}
\def \sem#1{\mbox{\lsem$#1$\rsem}}

\newcommand {\qc}[1] {{\sf{#1}}}
\def\>{\ensuremath{\rangle}}
\def\<{\ensuremath{\langle}}
\def\ott{t}
\def\otu{u}

\def\ctp{P}
\def\ctq{Q}

\def\dmu{\mu}
\def\dnu{\nu}

\def\qv{\mathit{qVar}}
\def\fv{\mathit{fv}}
\def\cv{\mathit{cVar}}
\def\be{\mathit{BExp}}
\def\ex{\mathit{Exp}}
\def\cc{\mathit{cChan}}
\def\qnc{\mathit{qChan}}
\def\ch{\mathit{Chan}}
\def\con{\mathit{Con}}
\def\Wait{\mathit{Wait}}
\def\so{\ensuremath{SO}}

\newcommand{\pdist}[1]{\overline{#1}  } 
\newcommand{\Act}{\ensuremath{\mathsf{Act}}}
\newcommand{\support}[1]{\lceil{#1}\rceil}
\newcommand{\setof}[2]{\{ \, #1 \, \mid \, #2 \, \}}

\def\h{\ensuremath{\mathcal{H}}}
\def\p{\ensuremath{\mathcal{P}}}
\def\dh{\ensuremath{\mathcal{D(H})}}
\def\r{\ensuremath{\mathcal{R}}}
\def\t{\ensuremath{\mathcal{T}}}
\def\ra{\ensuremath{\rightarrow}}
\def\e{\ensuremath{\mathcal{E}}}

\def\c{\ensuremath{\mathcal{C}}}
\def\d{\ensuremath{\mathcal{D}}}
\def\eval{\ensuremath{{\psi}}}
\def\dist{\ensuremath{D}}
\def\sdist{\ensuremath{D}}

\newcommand{\supp}[1]{\ensuremath{\lceil{#1}\rceil}}
\newcommand{\abis}{\stackrel{\lambda}\approx}
\newcommand{\abisa}[1]{\stackrel{#1}\approx}
\newcommand {\nil} {\mbox{\bf{nil}}}
\newcommand {\iif} {\mbox{\bf{if}}}
\newcommand {\then} {\mbox{\bf{then}}}
\newcommand {\eelse} {\mbox{\bf{else}}}
\newcommand {\true} {\rm{tt}}
\newcommand {\false} {\rm{{ff}}}
\newcommand{\tr}{{\rm tr}}
\newcommand{\rto}[1]{\stackrel{#1}\longrightarrow}
\newcommand{\srto}[1]{\stackrel{#1}\longrightarrow}
\newcommand{\Rto}[1]{\stackrel{#1}\Longrightarrow}

\newcommand{\define}{:=}
\newcommand{\env}{\mathrm{env}}

\title{Toward automatic verification of quantum cryptographic protocols}

\author[1,3]{Yuan Feng\footnote{Email address: \texttt{Yuan.Feng@uts.edu.au}}}
\author[1,2]{Mingsheng Ying}
\affil[1]{University of Technology Sydney, Australia}  
\affil[2]{Department of Computer Science and Technology,
National Laboratory for Information Science and Technology,
Tsinghua University, China}
\affil[3]{ AMSS-UTS Joint Research Laboratory for Quantum Computation, Chinese Academy of Sciences}
\authorrunning{Y. Feng and M. Ying} %mandatory. First: Use abbreviated first/middle names. Second (only in severe cases): Use first author plus 'et. al.'

\Copyright{Yuan Feng and Mingsheng Ying}%mandatory, please use full first names. LIPIcs license is "CC-BY";  http://creativecommons.org/licenses/by/3.0/


% mandatory: Please provide 1-5 keywords

\begin{document}

\maketitle

\begin{abstract}
 Several quantum process algebras have been proposed and successfully applied in verification of quantum cryptographic protocols. All of the bisimulations proposed so far for quantum processes in these process algebras are state-based, implying that they only compare individual quantum states, but not a combination of them. 
This paper remedies this problem by introducing a novel notion of distribution-based bisimulation for quantum processes. We further propose an approximate version of this bisimulation that enables us to prove more sophisticated security properties of quantum protocols which cannot be verified using the previous  bisimulations. In particular, we prove that the quantum key distribution protocol BB84 is sound and (asymptotically) secure against the intercept-resend attacks by showing that the BB84 protocol, when executed with such an attacker concurrently, is approximately bisimilar to an ideal protocol, whose soundness and security are obviously guaranteed, with at most an exponentially decreasing gap.  

 \end{abstract}

\section{Introduction}

Quantum cryptography can provide unconditional security; it allows the realisation of cryptographic tasks that are
proven or conjectured to be impossible in classical cryptography. The security of quantum cryptographic protocols
is mathematically provable, based on the principles of quantum mechanics, without imposing any restrictions on the
computational capacity of attackers. The proof is, however, often notoriously difficult,
which is evidenced by the 50 pages long security proof of the quantum key distribution protocol BB84~\cite{Mayers:2001bx}. 
It is hard to imagine such an analysis being carried out
for more sophisticated quantum protocols. Thus, techniques for (semi-)automated verification of quantum protocols will be indispensable, given that quantum communication systems are already commercially available. 

Process algebra has been successfully applied in the verification of classical (non-quantum) cryptographic protocols~\cite{Mitchell:2000da,Ramanathan:2004kx}.
One key step for such a process algebraic approach is a suitable notion of \emph{bisimulation} which has appropriate distinguishing power and is preserved by various process constructs. Intuitively,
two systems are bisimilar if and only if each observable action of one of them can be simulated by the other by performing the same observable action (possibly preceded and/or followed by some unobservable internal actions), and furthermore, the resultant systems are again bisimilar. To verify a cryptographic protocol, we first give a \emph{specification} which is an ideal protocol with obvious correctness and security, and then show that the given protocol is bisimilar (or approximately bisimilar with a small perturbation) to the specification. 

In the last 10 years, several quantum process algebras like CQP~\cite{GN05}, QPAlg~\cite{JL04} and qCCS~\cite{FDJY07} have been introduced, which provide an intuitive but rigorous way to model and reason about quantum communication systems. In particular, they have been adopted in verification of several popular quantum communication protocols such as Teleportation, Superdense Coding, etc. 
Similar to the classical case, the notion of bisimulation is crucial in the process algebra-based verification of quantum protocols. Actually, several different versions of bisimulation have been proposed for quantum processes in the recent literature~\cite{La06,YFDJ09,FDY11,Da11,DF11}. A key feature of all of them is that they are state-based in the sense that they only compare individual configurations but not a combination of them. 
More explicitly, they are defined to be relations over configurations which are pairs of a quantum process and a density operator describing the state of environment quantum systems. However, when distributions of configurations are considered (which is inevitable for protocols where randomness is employed or quantum measurement is involved), state-based bisimulations are too discriminative -- they distinguish some distributions which will never be distinguished by any outside observers, thereby providing the potential attacker of a cryptographic protocol with unrealistic power. 
As an extreme example, a state-based bisimulation distinguishes the distribution 
$p  \<\nil,  \rho\> + (1-p) \<\nil, \sigma\>$ from the single configuration $\<\nil, p \rho + (1-p)\sigma\>$ if $\rho\neq \sigma$, where $\nil$ is the \emph{dead} process incapable of performing any action.

In this paper, we propose a novel bisimulation for quantum processes which is defined directly on distributions of quantum configurations. Compared with existing bisimulations in the literature, our definition is strictly coarser (in particular, equates the two distributions presented above) and takes into account the combination of accompanied quantum states. 
We further  define a pseudo-metric to characterise the extent to which two quantum processes are bisimilar. Note that we only consider quantum processes written in qCCS, but the main results can be generalised to other quantum process algebras like CQP and QPAlg easily.

To illustrate the utility of distribution-based bisimulation and the pseudo-metric in verification of quantum cryptographic protocols, we analyse the soundness and security of the well-known BB84 quantum key distribution protocol~\cite{BB84}. For the soundness, we show that when executed alone (without the presence of an attacker), BB84 is bisimilar to an ideal protocol which always returns a uniformly distributed (conditioning on a given key size) key. For the security analysis, we prove that when BB84 is executed concurrently with an intercept-resend attacker, the whole system is approximately bisimilar, with at most an exponentially decreasing gap, to an ideal protocol which never reports failure or information leakage.
To the best of our knowledge, this is the first time (a weak notion of) {security} of BB84 is formally described and verified in the quantum process algebra approach.
 
%The remainder of the paper is organised as follows. We conclude this section by presenting a brief discussion on the related works. In Section~2, we review some
%basic definitions of probabilistic labelled transition systems, which serve as the semantic model of qCCS. Particularly, the technique of lifting a binary relation on states (or from states to distributions) to a relation on distributions of states is introduced. The syntax and transitional semantics of qCCS, as well as the state-based bisimulation proposed in~\cite{DF11}, are presented in Section~3. 
%Sections~4 and 5 are the main part of this paper where in the former we define a distribution-based bisimulation for qCCS processes, and show that it is  coarser than the bisimulation in~\cite{DF11}. Section 5 introduces the notion of distribution-based approximate bisimulation, which gives rise to a pseudo-metric between quantum processes.
%Section~6 is devoted to the soundness and security analysis of the BB84 protocol. 
% We outline the main results in Section 7 and point out some directions for further study. 
 
\emph{Related works}. The problem of existing bisimulations, as pointed out in the third paragraph of this section, was also noted by Kubota et al.~\cite{kubota2012application}. To deal with it, they adopted two different semantics for quantum measurements.
When a measurement induces a probability distribution in which 
all configurations have the same observable actions, it is represented semantically as a super-operator  obtained by discarding the measurement outcome (thus no probabilistic branching is produced, and all post-measurement quantum states are merged). Otherwise, the measurement has the same semantics as in the original qCCS. This treatment solves the problem when probabilistic behaviours are only induced by quantum measurements. However, it does not work when probabilistic choice is included in the syntax level, as we do in describing BB84 protocol in this paper. Furthermore, it brings difficulty in deciding the right semantics of a quantum process where a measurement is involved, as determining if the observable actions of the post-measurement configurations are all the same might not be easy; sometimes it even depends on the later input from the environment. In this paper, we solve this problem by revising the definition of bisimulation, instead of the definition of semantics.

In the same paper~\cite{kubota2012application},  Kubota et al. applied qCCS (with the semantic modification mentioned above) to show the security of BB84. They proved that BB84 is bisimilar to an EDP-based protocol, following the proof of Shor and Preskill~\cite{shor2000simple}. However, this should not be regarded as a complete security proof, as it relies on the security of the  EDP-based protocol. In contrast, our approach shows the security of BB84 directly. Note that for this purpose, a notion of approximate bisimulation, which was not presented in~\cite{kubota2012application}, is necessary, as BB84 is secure only in the sense that the eavesdropper's information about the secure key obtained by the legitimate parties is arbitrarily small (but still can be strictly positive) when the number of qubits transmitted (called the \emph{security parameter}) goes to infinity. 

Software tools based on the quantum process algebra CQP have been developed in~\cite{ardeshir2013equivalence} and~\cite{ardeshir2014verification} to check the equivalence between quantum sequential programs as well as concurrent protocols. These tools were applied to verify the correctness of protocols like Teleportation, Bit Flip Error Correction Code, and Quantum Secret Sharing. However, verification of security properties in cryptographic protocols such as BB84 has not been reported yet.

Besides the process algebra approach, model-checking is another promising approach for verification of quantum cryptographic protocols. For example, 
by observing the fact that the quantum states appearing in BB84, when only intercept-resend eavesdroppers are considered, are all the so-called stabiliser states which can be efficiently encoded in a classical way, Nagarajan et al.~\cite{nagarajan2005automated} analysed 
the security of BB84 by 
using the probabilistic model checker PRISM~\cite{Kwiatkowska:2004bo}.

\section{Preliminaries}
\label{sec:plts}

In this section we review the model of probabilistic labelled
transition systems (pLTSs) and the notion of lifted relations. Later on we will interpret the behaviour
of quantum processes in terms of pLTSs.
\subsection{Probabilistic labelled transition systems}
\label{subsec:plts}

A (finite-support) \emph{probability distribution}
over a set $S$ is a function $\mu : S \rightarrow [0, 1] $ with $\mu(s)>0$ for finitely many $s\in S$ and
$\sum_{s\in S} \mu(s) = 1$; the support of such a $\mu$ is
the set $\support{\mu} = \setof{s \in S}{\mu(s) > 0}$.  
The \emph{point distribution} $\pdist{s}$ assigns probability
$1$ to $s$ and $0$ to all other elements of $S$, so that
$\support{\pdist{s}} = \{s\}$. We use $\dist(S)$ to denote the set of probability distributions over $S$, ranged over by $\mu,\nu$ etc.
If $\sum_{i \in I} p_i = 1$ for some
collection of  $p_i \geq 0$, and $\mu_i\in \dist(S)$,
then $\sum_{i \in I}p_i \cdot \mu_i\in \dist(S)$ is a \emph{combined} probability distribution with
$(\sum_{i \in I}p_i \cdot \mu_i)(s)~=~\sum_{i\in I} p_i\cdot \mu_i(s).$
We always assume the index set $I$ to be finite.

\begin{definition}\label{def:LTS}
  A \emph{probabilistic labelled transition system} (pLTS) is a triple
$\langle S, \Act,  \rto{}  \rangle$, where
 $S$ is a set of \emph{states}, $\Act$ is a set of \emph{transition labels} with a special element $\tau$ included, and the \emph{transition relation} $\rto{}$ is a subset of
$S \times \Act  \times \dist(S)$.
\end{definition}

\subsection{Lifting relations}

In a pLTS actions are only performed by states, in that they are
given by relations from states to distributions. But in general we
allow distributions over states to perform an action. For this
purpose, we \emph{lift} these relations to
distributions \cite{DGHM09,DF11}. 
\begin{definition}[Lifting]\label{def:lift}
Let  $\mathord{\aRel} \subseteq
  S\times\sdist(S)$ be a relation.
The lifted relation, denoted by $\aRel$ again for simplicity, is the smallest relation $\aRel \subseteq \sdist(S) \times
\sdist(S)$ that satisfies
\begin{enumerate}
\item $s \aRel \nu$ implies $\pdist{s} \aRel \nu$, and
\item\label{i1529} (Linearity)
 $\mu_i \aRel \nu_i$ for $i\in I$ implies
 $(\sum_{i\in I}p_i\cdot\mu_i)\aRel(\sum_{i\in I}p_i\cdot\nu_i)$
 for any $p_i \in [0,1]$ with $\sum_{i\in I}p_i = 1$.
\end{enumerate}
\end{definition}

We apply this operation to the relations  $\ar{\alpha}$ in a pLTS
for $\alpha\in \Act$. Thus as source of
a relation $\ar{\alpha}$ we also allow distributions.
But  $\pdist{s} \ar{\alpha} \mu$ is more general than 
$s \ar{\alpha} \mu$, because if
$\pdist{s}\ar{\alpha}\mu$ then there is a collection of distributions
$\mu_i$ and probabilities $p_i$ such that $s\ar{\alpha}\mu_i$ for each $i\in I$ and
$\mu=\sum_{i\in I}p_i\cdot\mu_i$ with $\sum_{i\in I}p_i=1$; that is, we allow different transitions to be combined together, provided that 
they have the same source $s$ and the same label $\alpha$. 

Sometimes we also need to lift a relation on states, say a state-based bisimulation, to distributions. This can be done by the following two steps. Let $\aRel\subseteq S \times S$ be such a relation. First, it induces a relation $\hat{\aRel}\subseteq S\times\dist(S)$ between states and distributions:
$\hat{\aRel} \Defs \sset{(s,\pdist{t})\mid s\aRel t}.$
Then we can use Definition~\ref{def:lift} to lift $\hat{\aRel}$ to distributions. Note that when $\aRel$ is an equivalence relation over $S$, the lifted relation  over $\dist(S)$ coincides with the lifting defined in~\cite{jonsson2001probabilistic}.

In Definition~\ref{def:lift}, linearity tells us how to compare two
linear combinations of distributions. Sometimes we need a dual
notion of decomposition. Intuitively, if a relation $\aRel$ is
\emph{left-decomposable} and $\mu \aRel \nu$, then for any
decomposition of $\mu$ there exists some corresponding
decomposition of $\nu$.
\begin{definition}[Left-decomposable]\label{def:left-deconst}
 A binary relation over distributions,  $\mathord{\aRel} \subseteq
  \dist(S) \times\dist(S)$,
 is called \emph{left-decomposable} if
 $(\sum_{i\in I}p_i\cdot\mu_i)   \aRel \nu$
implies that
   $\nu$ can be written as $(\sum_{i\in I}p_i\cdot\nu_i)$
    such that $\mu_i \aRel \nu_i$ for every $i \in I$.
\end{definition}

The next lemma shows that any lifted relation is left-decomposable.
\begin{lemma}[\cite{DF11}]\label{lem:leftdec}
  For any $\aRel \subseteq S \times \dist(S)$ or $S\times S$, the lifted relation 
 over distributions is left-decomposable. 
\end{lemma}

With the help of lifted relations, we are now able to define various (weak) transitions between distributions for a pLTS.
\begin{definition} Given a pLTS $\langle S, \Act,  \rto{}  \rangle$, we define the following transitions over distributions:
\begin{enumerate}
\item $\ar{\hat{\tau}}$. Let $s \ar{\hat{\tau}} \mu$ if either $s \ar{\tau}\mu$ or $\mu = \pdist{s}$, and lift it to distributions;
\item $\ar{\hat{\alpha}}$ for $\alpha\neq \tau$. Let $s \ar{\hat{\alpha}} \mu$ if $s\ar{\alpha}\mu$, and lift it to distributions;
\item $\dar{\hat{\tau}}$.  Let ${\dar{\hat{\tau}}} = {(\ar{\hat{\tau}})^*}$ be the
reflexive and transitive closure of $\ar{\hat{\tau}}$;
\item $\dar{\hat{\alpha}}$ for $\alpha\neq \tau$. Let ${\dar{\hat{\alpha}}} ={
\dar{\hat{\tau}} \ar{\hat{\alpha}} \dar{\hat{\tau}}}$. For point distributions, we often write 
$s\dar{\hat{\alpha}}\nu$ instead of $\ \pdist{s}\dar{\hat{\alpha}}\nu$.
\end{enumerate}
\end{definition}

Note that here $\dar{\hat{\alpha}}$ is not a lifted transition. However, the next lemma shows that it is still both linear and left-decomposable.
\begin{lemma}[\cite{DF11}]
 The transition relations $\dar{\hat{\alpha}}$ are both linear and 
left-decomposable.
\end{lemma}

\section{qCCS: Syntax and Semantics}
In this section, we review the syntax and semantics of qCCS, a quantum extension of value-passing CCS 
introduced in \cite{FDJY07,YFDJ09}, and a notion of state-based bisimulation for qCCS processes presented in~\cite{DF11}.
We assume the readers are familiar with the basic notions
in quantum information theory; for those who are not, please refer to \cite{NC00}.

\subsection{Syntax}

We assume three types of data in qCCS: \qc{Bool} for booleans,  real numbers \qc {Real}
for classical data, and qubits \qc {Qbt} for quantum data. Let $\cv$, ranged over
by $x,y,\dots$, be the set of classical variables, and $\qv$, ranged over by $q,r,\dots$, the set of
 quantum variables. It is assumed that $\cv$ and $\qv$ are both countably infinite.
 We assume a set $\ex$, which includes $\cv$ as a subset and is ranged over by $e,e',\dots$,  of classical expressions over
\qc {Real}, and a set of boolean-valued expressions $\be$, ranged over by $b, b',\dots$, with the usual set of boolean operators $\true$, $\false$,
$\neg$, $\wedge$, $\vee$, and $\ra$. In particular, we let $e\bowtie e'$ be a boolean expression for any $e,e'\in \ex$ and ${\bowtie} \in \{ >, <, \geq, \leq, =\}$.
We further assume that only classical variables can occur free in data expressions and boolean expressions.
 Let $\cc$
be the set of classical channel names, ranged over by $c,d,\dots$,
and $\qnc$ the set of quantum channel names, ranged over by $\qc
c,\qc d,\dots$. Let $\ch=\cc\cup \qnc$. A relabeling function
$f$ is a one-to-one function from $\ch$ to $\ch$ such that
$f(\cc)\subseteq \cc$ and $f(\qnc)\subseteq \qnc$.

We often abbreviate the indexed set
$\{q_1,\dots,q_n\}$ to $\tilde{q}$ when $q_1, \dots,q_n$ are
distinct quantum variables and the dimension $n$ is understood. Sometimes we also use $\tilde{q}$ to denote
the string $q_1\dots q_n$. 
We assume a set of process constant schemes, ranged over by 
$A, B, \dots$. Assigned to each process constant scheme $A$ there are two non-negative 
integers $ar_c(A)$ and $ar_q(A)$. If $\tilde{x}$ is a tuple of classical variables with
$|\tilde{x}|=ar_c(A)$, and $\tilde{q}$ a tuple of distinct quantum variables with
$|\tilde{q}|=ar_q(A)$, then $A(\tilde{x},\tilde{q})$ is
called a process constant. When $ar_c(A)=ar_q(A)=0$, we also 
denote by $A$ the (unique) process constant produced by $A$.

The syntax of qCCS terms can be given by the Backus-Naur form as
\begin{eqnarray*}
\ott &::=& \nil\ |\ A(\tilde{e}, \tilde{q})\ |\ \alpha.\ott\ |\ \ott+\ott\ |\ \ott\| \ott\ |\ \ott\backslash L\ |\ \ott[f]\ |\ \iif \ b\ \then \ \ott\\
\alpha &::= &\tau\ |\ c?x\  |\ c!e\ |\ \qc c?q\ |\ \qc c!q \ |\ \e[\tilde{q}]\ |\ M[\tilde{q};x]
\end{eqnarray*}
where $c\in \cc$, $x\in \cv$, $\qc c\in \qnc$,
$q\in \qv$, $\tilde{q}\subseteq \qv$, $e\in \ex$, $\tilde{e}\subseteq \ex$, $\tau$ is the silent action,
$A$ is a
process constant scheme, $f$ is a relabeling function, $L\subseteq \ch$,
$b\in \be$, $\e$ and $M$ are respectively
a super-operator and a quantum measurement applying on the Hilbert
space associated with the systems $\tilde{q}$.

To exclude quantum processes which are not physically implementable, we also require $q\not\in qv(\ott)$ in $\qc c!q.\ott$ and
$qv(\ott)\cap qv(\otu)=\emptyset$ in $\ott \| \otu$, where for a process term $\ott$, $qv(\ott)$ is the set of its free quantum variables which are not bound
by quantum input $\qc c?q$.
The notion of
free classical variables in quantum processes can be defined in the
usual way with the only modification that the quantum measurement prefix
$M[\tilde{q};x]$ has binding power on $x$. A quantum process term $\ott$
is closed if it contains no free classical variables, $i.e.$,
$\fv(\ott)=\emptyset$. 
We let $\t$, ranged over by $\ott, \otu, \cdots$, be the set of all qCCS terms, and $\p$, ranged over by $\ctp, \ctq, \cdots$, the set of closed terms.
To complete the definition of qCCS syntax, we assume that for each process constant 
$A(\tilde{x}, \tilde{q})$, there is a defining equation 
$A(\tilde{x}, \tilde{q})\define \ott$
where $\fv(\ott)\subseteq \tilde{x}$ and $qv(\ott)\subseteq \tilde{q}$. Throughout the paper we implicitly assume the convention that process terms are identified up to $\alpha$-conversion.

The process constructs we give here are quite similar to those in
classical CCS, and they also have similar intuitive meanings: $\nil$
stands for a process which does not perform any action; $c?x$ and $
c!e$ are respectively classical input and classical output, while
$\qc c?q$ and $\qc c!q$ are their quantum counterparts. $\e[\tilde{q}]$
denotes the action of performing the quantum operation $\e$ on the
qubits $\tilde{q}$ while $M[\tilde{q};x]$ measures the qubits $\tilde{q}$
according to $M$ and stores the measurement outcome into the
classical variable $x$. $+$ models nondeterministic choice: $\ott+\otu$
behaves like either $\ott$ or $\otu$ depending on the choice of the
environment. $\|$ denotes the usual parallel composition. The
operators $\backslash L$ and $[f]$ model restriction and relabeling,
respectively: $\ott\backslash L$ behaves like $\ott$ but any action
through the channels in $L$ is forbidden, and $\ott[f]$ behaves like
$\ott$ where each channel name is replaced by its image under the
relabeling function $f$. Finally, $\iif\ b\ \then\ \ott$ is the
standard conditional choice where $\ott$ can be executed only if $b$ evaluates to
\true.

An evaluation $\eval$ is a function from $\cv$ to \qc{Real}; it can be extended in an obvious way to functions from $\ex$ to $\qc{Real}$ and from $\be$ to $\{\true, \false\}$, and finally, from $\t$ to $\p$. For simplicity, we still use $\eval$ to denote these extensions. Let $\eval\{v/x\}$ be the evaluation which differs from $\eval$ only in that it maps $x$ to $v$.

\subsection{Transitional semantics}

For each quantum variable $q\in  \qv$, we assume a 2-dimensional 
Hilbert space $\h_q$ to be the
state space of the $q$-system. For any $V\subseteq \qv$,  we denote
$\h_{V}=\bigotimes_{q\in V} \h_q.$
In particular, $\h = \h_{\qv}$ is the state space of the whole environment consisting of
all the quantum variables. Note
that $\h$ is a countably-infinite dimensional Hilbert space.
For any $V\subseteq \qv$ we denote by $\overline{V}$ the 
complement set of $V$ in $\qv$.

Suppose $\ctp$ is a closed quantum process. A pair of the form
$\<\ctp,\rho\>$ is called a configuration, where $\rho\in \dh$ is a density operator
on $\h$.\footnote{As $\h$ is infinite dimensional, $\rho$ should be understood as a density operator on some finite dimensional subspace of $\h$ which contains $\h_{qv(P)}$.} The set of configurations is denoted by $\con$, and ranged over by $\c,\d,\dots$. 
Let
\begin{eqnarray*}
\Act&=&\{\tau\}\cup\{c?v,c!v\ |\ c\in \cc, v\in \qc{Real}\}\cup\{\qc c?r,\qc c!r\ |\ \qc c\in \qnc, r\in \qv\}.
\end{eqnarray*}
For each $\alpha\in \Act$, we define the bound quantum variables $qbv(\alpha)$ of $\alpha$ as 
$qbv(\qc c?r) = \{r\}$ and $qbv(\alpha)=\emptyset$ if $\alpha$ is not a quantum input.
The  channel names used in action $\alpha$ is denoted by $cn(\alpha)$; 
that is, $cn(c?v) = cn(c!v) = \{c\}$, $cn(\qc c?r) = cn(\qc c!r) = \{\qc c\}$, and $cn(\tau)=\emptyset$. We also extend the relabelling function
to $\Act$ in an obvious way.
Then the transitional semantics of qCCS can be
given by a pLTS
$\<\con,\Act,\srto{}\>$, where ${\srto{}}\subseteq \con\times
\Act\times \dist(\con)$ is the smallest relation satisfying the inference rules
depicted in Fig.~\ref{fig:csem}. 
The symmetric forms for rules $Par$, $Com_C$, $Com_Q$, and $Sum$ are omitted.
We abuse the notation slightly by writing $\c\rto{{\alpha}}\d$ if $\c\rto{{\alpha}}\pdist{\d}$. We also use the obvious extension of the function $\|$ on configurations to distributions. To be precise,
if $\mu=\sum_{i\in I}p_i \<\ctp_i, \rho_i\>$ then $\mu\| \ctq$ denotes the distribution
$\sum_{i\in I}p_i \<\ctp_i\| \ctq, \rho_i\>$.
Similar extension applies to $\dmu[f]$ and $\dmu\backslash L$.

\begin{figure*}[t] {
\scriptsize
\[
\begin{array}{rlrl}
\mathit{Tau} & \displaystyle \frac{}{
\<\tau.\ctp,\rho\> \srto{\tau} {\<\ctp,\rho\>}}&
\mathit{Inp_C} & \displaystyle \frac{v\in \qc
{Real}}{
\<c?x.\ott,\rho\> \srto{c?v} {\<\ott\{v/x\},\rho\>}}\\
\\
\mathit{Out_C}& \displaystyle \frac{v=\sem{e}}{
\<c!e.\ctp,\rho\> \srto{c!v} {\<\ctp,\rho\>}} &
\mathit{Inp_Q} & \displaystyle \frac{r\not\in qv(\qc c?q.\ctp)}{ \<\qc c?q.\ctp,\rho\>
\srto{\qc c?r} {\<\ctp\{r/q\},\rho\>}}\\
\\
\mathit{Out_Q} & \displaystyle \frac{}{ \<\qc c!q.\ctp,\rho\>
\srto{\qc c!q} {\<\ctp,\rho\>}}
&
\mathit{Oper}&
\displaystyle \frac{}{ \<\e[\tilde{r}].\ctp ,\rho\>
\srto{\tau} {\<\ctp,\e_{\tilde{r}}(\rho)\>}}\\
\\
\mathit{Meas} & 
\displaystyle \frac{M=\sum_{i\in I} \lambda_i E^i,\ p_i=\tr(E^i_{\tilde{r}}\rho)>0}{
\<M[\tilde{r};x].\ctp ,\rho\> \srto{\tau}\sum_{i\in I}
p_i \<\ctp\{\lambda_i/x\},E^i_{\tilde{r}}\rho
 E^i_{\tilde{r}}/p_i\>} &
\mathit{Par} & \displaystyle \frac{
\<\ctp_1,\rho\>\rto{{\alpha}} \dmu,\ qbv(\alpha)\cap qv(\ctp_2)=\emptyset}{ \<\ctp_1\|\ctp_2,\rho\>\rto{{\alpha}}
\dmu\|\ctp_2}\\
\\
\mathit{Com_C} & \displaystyle \frac{ \<\ctp_1,\rho\>\srto{c?v}
{\<\ctp_1',\rho\>},\ \<\ctp_2,\rho\>\srto{c!v} {\<\ctp_2',\rho\>}}
{ \<\ctp_1\|\ctp_2,\rho\>\srto{\tau}
{\<\ctp_1'\|\ctp_2',\rho\>}}&
\mathit{Com_Q}& \displaystyle \frac{ \<\ctp_1,\rho\>\srto{\qc
c?r} {\<\ctp_1',\rho\>},\hspace{1em} \<\ctp_2,\rho\>\srto{\qc c!r}
{\<\ctp_2',\rho\>}}{ \<\ctp_1\|\ctp_2,\rho\>\srto{\tau}
\<\ctp_1'\|\ctp_2',\rho\>}\\
\\
\mathit{Sum} & \displaystyle \frac{ \<\ctp,\rho\>\rto{{\alpha}}
\dmu}{ \<\ctp+\ctq,\rho\>\rto{{\alpha}} \dmu} &
\mathit{Rel} & \displaystyle \frac{
\<\ctp,\rho\>\rto{{\alpha}}\dmu}{
\<\ctp[f],\rho\>\srto{f(\alpha)} \dmu[f]}
\\ \\
\mathit{Cho} & \displaystyle \frac{
\<\ctp,\rho\>\rto{{\alpha}}\dmu,\ \sem{b}=\mbox{\true}}{ \<\iif\ b\ \then\ \ctp,\rho\>\rto{{\alpha}}\dmu}
&
\mathit{Res} &
 \displaystyle \frac{ \<\ctp,\rho\>\rto{{\alpha}} \dmu,\ cn(\alpha)\cap L=\emptyset}{ \<\ctp\backslash L,\rho\>\rto{{\alpha}} \dmu\backslash L}
\\
\\
\mathit{Def} & \displaystyle \frac{
\<\ott\{\tilde{v}/\tilde{x},\tilde{r}/\tilde{q}\},\rho\>\rto{{\alpha}}\dmu,\ A(\tilde{x},\tilde{q})\define \ott,\ \tilde{v}=\sem{\tilde{e}}}{
\<A(\tilde{e},\tilde{r}),\rho\>\rto{{\alpha}}\dmu}& &
\end{array}
\]}
 \caption{Transitional semantics of qCCS \label{fig:csem}}
\end{figure*}

\subsection{State-based bisimulation}

In this subsection, we recall the basic definitions and properties of the state-based bisimulation introduced in~\cite{DF11}.
Let $\c=\<P,\rho\>$ be a configuration and $\e$ a super-operator. We denote $qv(\c)= qv(P)$, $\env(\c)=\tr_{qv(P)}(\rho)$ being the quantum \emph{environment} of process $P$ in $\c$, and $\e(\c) = \<P, \e(\rho)\>$. Furthermore, for distribution $\mu = \sum_i p_i \c_i$ with $p_i>0$ for each $i$, we write $qv(\mu)= \bigcup_i qv(\c_i)$, $\env(\mu)= \sum_i p_i \cdot \env(\c_i)$, and $\e(\mu) = \sum_i p_i \e(\c_i)$. For any $V\subseteq \qv$, denote by $\so(\h_V)$ the set of super-operators on $\h_V$.

\begin{definition}
A relation $\aRel \subseteq \con\times\con$ is closed under super-operator application if $\c\aRel\d$ implies $\e(\c)\aRel \e(\d)$ for all $\e\in \so(\h_{\overline{qv(\c)\cup qv(\d)}})$. More generally, a relation $\aRel \subseteq \sdist(\con)\times\sdist(\con)$ is closed under super-operator application if $\mu\aRel\nu$ implies $\e(\mu)\aRel \e(\nu)$ for all $\e\in \so(\h_{\overline{qv(\mu)\cup qv(\nu)}})$.
\end{definition}

\begin{definition}\label{def:sbisimulation}
\begin{enumerate}
\item A symmetric relation $\r\subseteq \con\times \con$ is called a
\emph{state-based ground bisimulation} if $\c\r \d$ implies that 
\begin{enumerate}
\item[(i)] $qv(\c)=qv(\d)$, and $\env(\c) = \env(\d)$, 
\item[(ii)] whenever $\c \rto{\alpha} \dmu$, there
exists $\dnu$ such that $\d\Rto{\hat{\alpha}}\dnu$ and
$\dmu\r\dnu$.
\end{enumerate}
\item 
A relation $\aRel$ is a \emph{state-based bisimulation} if it is a state-based ground bisimulation, and is closed under super-operator application.
\item Two quantum configurations $\c$ and $\d$ are state-based bisimilar, denoted by
$\c\approx_s \d$, if there exists a state-based bisimulation $\r$ such that
$\c\r \d$;
\item
Two quantum process terms $\ott$ and $\otu$ are state-based bisimilar, denoted by
$\ott\approx_s \otu$, if for any quantum state $\rho\in \d(\h)$ and any evaluation $\eval$, $\< \ott\eval, \rho\>\approx_s \<\otu\eval, \rho \>.$ 
\end{enumerate}
\end{definition}

Note that in Clause 1.(ii) of the above definition, $\dmu\r\dnu$ means $\mu$ and $\nu$ are related by the relation lifted from $\aRel$. 
The following theorem is taken from~\cite{DF11}.

\begin{theorem} \label{th:propsbis}
\begin{enumerate}
\item The bisimilarity relation $\approx_s$ is the largest state-based bisimulation on $\con$, and it is an equivalence relation.
\item  As a lifted relation on $\dist(\con)$,  $\approx_s$ is both linear and left-decomposable.
\end{enumerate}
\end{theorem}

\section{Distribution-based bisimulation}

Note that in \cite{Doyen:2008uq}, it has already been shown by examples that state-based bisimulation is sometimes too discriminative for probabilistic automata. 
These examples certainly work for quantum processes as well. Furthermore, as the following example indicates, the problem becomes more serious in the quantum setting, as the accompanied quantum states can and should be combined when simulating each other.
\begin{example}\label{exam:counter}
Let $M = \lambda_0|0\>\<0| + \lambda_1|1\>\<1| $ be a two-outcome measurement according to the computational basis, and $\e$ a super-operator with the Kraus operators being $|0\>\<0|$ and $|1\>\<1|$. 
Let $\rho$ be a density operator on $\h_{\overline{\{q\}}}$, and $\c :=  \<M[q;x].\nil, |+\>_q\<+| \otimes \rho\>$ and $\d :=  \<\e[q].\nil, |+\>_q\<+| \otimes \rho\>$ be two configurations where $|+\> = (|0\> + |1\>)/\sqrt{2}$.
Note that in the process $M[q;x].\nil$, the measurement outcome is never used (as $x\not\in \fv(\nil)$), while the effect of $\e[q]$ is exactly measuring the system $q$ according to $M$, but ignoring the measurement outcome. Thus we definitely would like to regard $\c$ and $\d$ as being bisimilar\footnote{Note that $\c$ and $\d$ would be regarded as `semantically identical' in~\cite{kubota2012application}, instead of `(distribution-based) bisimilar' as we do in this paper, since the semantics of $M[q;x]$ in this case is represented as $\e[q]$ \emph{by definition}.}.

However, we can show that $\c\not\approx_s \d$. 
Let $\c_0 = \<\nil, |0\>_q\<0| \otimes \rho\>$,  $\c_1 = \<\nil, |1\>_q\<1| \otimes \rho\>$,  $\c_I = \<\nil, I_q/2 \otimes \rho\>$, and $\mu = \frac 12 \c_0 + \frac 12 \c_1$. Then obviously $\mu\not\approx_s \pdist{\c_I}$, as otherwise by the left-decompositivity of $\approx_s$ we must have both $\c_0 \approx_s \c_I$ and $\c_1 \approx_s \c_I$, which is impossible.  
\end{example}

Actually, the argument in Example~\ref{exam:counter} applies to \emph{any} bisimulation which is state-based: by Lemma~\ref{lem:leftdec}, any bisimilation between distributions which is lifted from configurations is left-decomposable, hence discriminating $\c$ and $\d$. Therefore, to make these two obviously indistinguishable configurations bisimilar, we have to define bisimulation relation \emph{directly} on distributions, rather than on configurations and then lift it to distributions.

For this purpose, we extend the distribution-based bisimulation introduced in \cite{Eisentraut:2012wi} to our quantum setting. 
A distribution $\mu$ is said to be \emph{transition consistent}, if for any $\c\in \supp{\mu}$ and $\alpha\neq \tau$, $\c\Rto{\hat{\alpha}} \nu_\c$ for some $\nu_\c$ implies $\mu\Rto{\hat{\alpha}} \nu$ for some $\nu$, i.e., all configurations in its support have the same set of enabled visible actions (possibly after some invisible transitions). Furthermore, a decomposition $\mu = \sum_{i\in I} p_i \cdot \mu_i$, $p_i>0$ for each $i\in I$, is a \emph{tc-decomposition} of $\mu$ if for each $i\in I$, 
$\mu_i$ is transition consistent.
\begin{definition}\label{def:dbisimulation}
\begin{enumerate}
\item A symmetric relation $\r\subseteq \sdist(\con)\times \sdist(\con)$ is called a
\emph{(distribution-based) ground bisimulation} if for any $\mu, \nu\in \sdist(\con)$, $\mu\r \nu$ implies that 
\begin{enumerate}

\item[(i)] $qv(\dmu)=qv(\dnu)$, and $\env(\dmu) = \env(\dnu)$, 

\item[(ii)] whenever $\mu \rto{\hat{\alpha}} \dmu'$, there
exists $\dnu'$ such that $\dnu\Rto{\hat{\alpha}}\dnu'$ and
$\dmu'\r\dnu'$,

\item[(iii)] if $\mu$ is not transition consistent, and $\mu = \sum_{i\in I} p_i \cdot \mu_i$ is a tc-decomposition, then $\nu \Rto{\hat{\tau}}  \sum_{i\in I} p_i \cdot \nu_i$ such that for each $i$, $\mu_i \r\nu_i$.

\end{enumerate}
\item 
A relation $\aRel$ is a  \emph{(distribution-based)  bisimulation} if it is a  ground bisimulation, and is closed under super-operator application.
\end{enumerate}

\end{definition}

In contrast with Definition~\ref{def:sbisimulation}.1, the above definition has an additional requirement Clause 1.(iii). This clause is crucial
for distribution-based bisimulation, as the transition $\mu \rto{\hat{\alpha}} \dmu'$ in Clause 1.(ii) is possible
only when $\mu$ is transition consistent for $\alpha$. That is, all configurations in the support of $\mu$ can perform weak $\alpha$-transition.
For those actions for which $\mu$ is not transition consistent, we must first split $\mu$ into transition consistent components, and then compare them with the corresponding components of $\nu$ individually.

The bisimilarity $\approx$ for quantum configurations and for quantum process terms are defined similarly as in the state-based case. The next theorem collects some useful properties of the distribution-based bisimilarity.
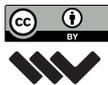
\begin{theorem} \label{thm:linearity}
\begin{enumerate}
\item The bisimilarity relation $\approx$ is the largest bisimulation on $\dist(\con)$, and it is an equivalence relation.
\item  $\approx$ is linear, but not left-decomposable.
\end{enumerate}
\end{theorem}

A direct consequence of Theorem~\ref{thm:linearity} is a deciding algorithm for the bisimilarity between recursion-free quantum configurations, which is sufficient for most practical quantum cryptographic protocols.
First, as pointed out in~\cite{FDY14}, any recursion-free quantum processes can be modified to be free of quantum input, so that the bisimilarity between them can be verified by only examining the ground bisimulation. Second, it has been proved in~\cite[Lemma 1]{HermannsKK14} that every linear
bisimulation $\r$ corresponds to a matrix $E$, so that two distributions $\mu$ and
$\nu$ are related by $\r$ if and only if $(\mu-\nu)E=0$, where distributions are
seen as vectors. As our ground bisimulation for quantum processes is indeed linear, the algorithm 
presented in~\cite{HermannsKK14}, with slight changes, can be used to decide it. For the sake of space limit, we omit the details here, and refer interested readers to~\cite{HermannsKK14}. 

To conclude this section, we would like to show that our distribution-based bisimulation is weaker than its state-based counterpart presented in Definition~\ref{def:sbisimulation}.
\begin{theorem}\label{thm:biscomp} Let $\mu, \nu\in \dist(\con)$. Then
 $\mu\approx_s \nu$ implies $\mu \approx \nu$, but $\mu\approx \nu$ does not necessarily imply $\mu \approx_s \nu$. In particular, we have in Example~\ref{exam:counter} that $\mu\approx \pdist{\c_I}$
 and $\c\approx \d$.
\end{theorem}

\section{Bisimulation metric}

In the previous section, only
\emph{exact} bisimulation is presented where two quantum processes are either bisimilar or non-bisimilar. Obviously, such a bisimulation cannot capture the idea that a quantum process \emph{approximately} implements its specification. 
To measure the behavioural distance between processes, the notion of approximate bisimulation and the bisimulation distance 
for qCCS processes were introduced in \cite{YFDJ09}. This section is devoted to extending this approximate bisimulation to distribution-based case.
Note that approximate bisimulation has been investigated in probabilistic process algebra and probabilistic labelled transition systems in the context of security analysis \cite{di2005measuring,aldini2008estimating}.

Recall that the trace distance of $\rho, \sigma\in \d(\h)$ is defined to be
$d(\rho, \sigma) = \frac{1}{2}\|\rho-\sigma\|_{\tr}$ where $\|\cdot\|_{\tr}$ denotes the trace norm. We have the following definition.

\begin{definition}\label{def:abis} 
Given $\lambda\in [0,1]$, a symmetric relation $\aRel$ over $\sdist(\con)$ which is closed under super-operator application is a $\lambda$-bisimulation if
for any $\mu \aRel \nu$, we have 
  \begin{enumerate}
\item $qv(\dmu)=qv(\dnu)$, and $d(\env(\dmu), \env(\dnu))\leq \lambda$,

\item whenever $\mu \rto{\hat{\alpha}} \dmu'$, there
exists $\dnu'$ such that $\dnu\Rto{\hat{\alpha}}\dnu'$ and
$\dmu'\aRel\dnu'$,

\item if $\mu$ is not transition consistent, and $\mu = \sum_{i\in I} p_i \cdot \mu_i$ is a tc-decomposition, then $\nu \Rto{\hat{\tau}}  \sum_{i\in I} p_i \cdot \nu_i$ such that $\sum_{i:\mu_i \aRel\nu_i} p_i \geq 1-\lambda$.
\end{enumerate}
\end{definition}

By induction, we can show easily that $\mu \rto{\hat{\alpha}} \dmu'$ can be replaced by $\mu \Rto{\hat{\alpha}} \dmu'$ in Clause (2). 

The approximate bisimilarity $\abis$ for quantum configurations and for quantum process terms are defined similarly as in the exact bisimulation case.
Furthermore, we define the bisimulation distance between distributions as
$
d_b(\mu, \nu) = \inf\{\lambda \geq 0 \mid \mu\abis \nu\}
$
and the bisimulation distance between process terms as
$d_b(\ott, \otu) = \inf\{\lambda \geq 0 \mid  \forall\eval \mbox{ and }\rho\in \d(\h), \< \ott\eval, \rho\>\abis \<\otu\eval, \rho \>\}.
$
Here we assume that $\inf\emptyset = 1$. The next theorem shows that $d_b$ is indeed a pseudo-metric with $\approx$ being its kernel.

\begin{theorem}\label{thm:pseudo}
\begin{enumerate}
\item
The bisimulation distance $d_{b}$ is a pseudo-metric on $\sdist(\con)$.
\item
For any $\mu, \nu\in \sdist(\con)$, $\mu \approx \nu$ if and only if $d_b(\mu, \nu) =0$.
\end{enumerate}
\end{theorem}
\section{An illustrative example}
\label{sec:examples}

For the ease of notations, we extend the syntax of qCCS a little bit by allowing probabilistic choice in the syntax level\footnote{Note that this extension will not change the expressive power of qCCS and all the results obtained in this paper, as probabilistic choices can be simulated by quantum measurements preceded by appropriate quantum state preparation.}; that is, we assume $\sum_{i\in I} p_i \ott_i\in\t$ whenever $\ott_i\in \t$ and $p_i \geq 0$ for  each $i\in I$ with $\sum_{i\in I} p_i =1$.
We further extend the transitional semantics in Fig.~\ref{fig:csem} by adding the following transition rule:
$$\mathit{Dist}\ 
  \frac{ }{ \<\sum_{i\in I} p_i \ott_i,\rho\>\srto{\tau} \sum_{i\in I} p_i \<\ott_i, \rho\>}.$$
We also introduce the syntax sugar $\iif\ b\ \then\ \ott\ \eelse\ \otu$ to be the abbreviation of 
$\iif\ b\ \then\ \ott\ +\ \iif\ \neg b\ \then\ \otu.$

BB84, the first quantum key distribution protocol developed by Bennett and Brassard in 1984~\cite{BB84}, provides a provably secure way to
create a private key between two parties, say, Alice and Bob, with the help of a classical authenticated channel and a quantum insecure channel between them.
Its security relies on the basic property of quantum mechanics that information gain about a quantum state is only possible at the expense of changing the state, if all the possible states are not orthogonal.
The basic BB84 protocol with security parameter $n$ goes as follows:
\begin{enumerate}
\item[(1)] Alice randomly generates two strings $\tilde{B}_a$ and $\tilde{K}_a$ of bits, each with size $n$.
\item[(2)] Alice prepares a string of qubits $\tilde{q}$, with size $n$, such that
the $i$th qubit of $\tilde{q}$ is $|x_y\>$ where $x$ and $y$ are the $i$th bits of $\tilde{B}_a$ and  $\tilde{K}_a$, respectively,
and $|0_0\> = |0\>$, $|0_1\> = |1\>$,
$|1_0\> = |+\>$, and $|1_1\> = |-\>$. Here $|+\> \define (|0\> + |1\>)/\sqrt{2} \mbox{ and }
|-\> \define (|0\> - |1\>)/\sqrt{2}. $
\item[(3)] Alice sends the qubit string $\tilde{q}$ to Bob.
\item[(4)] Bob randomly 
generates a string of bits $\tilde{B}_b$ with size $n$.
\item[(5)] Bob measures each qubit received from Alice according to a basis determined by the bits he generated: if the $i$th bit of $\tilde{B}_b$ is $k$ then
he measures with $\{|k_0\>, |k_1\>\}$, $k=0,1$. Let the measurement results be $\tilde{K}_b$, again a string of bits with size $n$.
\item[(6)] Bob sends his measurement bases $\tilde{B}_b$ back to Alice, and upon receiving the information, Alice sends her bases
$\tilde{B}_a$ to Bob.
\item[(7)] Alice and Bob determine at which positions the bit strings $\tilde{B}_a$ and  $\tilde{B}_b$ are equal. They discard the bits in
$\tilde{K}_a$ and $\tilde{K}_b$ where the corresponding bits of $\tilde{B}_a$ and  $\tilde{B}_b$ do not match.
\end{enumerate}
After the execution of the basic BB84 protocol above, the remaining bits of $\tilde{K}_a$ and $\tilde{K}_b$, denoted by $\tilde{K}'_a$ and $\tilde{K}'_b$ respectively, should be the same, provided that the channels used are perfect, and no eavesdropper exists. 

To detect a potential eavesdropper Eve, Alice and Bob proceed as follows:
\begin{enumerate}
\item[(8)] Alice randomly chooses $\supp{|\tilde{K}'_a|/2}$ bits of $\tilde{K}'_a$, denoted by $\tilde{K}''_a$, and sends to
Bob $\tilde{K}''_a$ and its indexes in $\tilde{K}'_a$.

\item[(9)]  Upon receiving the information from Alice, Bob sends back to Alice his substring $\tilde{K}''_b$ of $\tilde{K}'_b$ at the indexes received from Alice.

\item[(10)] Alice and Bob check if the strings $\tilde{K}''_a$ and $\tilde{K}''_b$ are equal. If yes, then the remaining substring $\tilde{K}^f_a$ (resp. $\tilde{K}^f_b$) of $\tilde{K}'_a$ (resp. $\tilde{K}'_b$) by deleting $\tilde{K}''_a$ (resp. $\tilde{K}''_b$) is the secure key shared by Alice (reps. Bob). Otherwise, an eavesdropper (or too much noise in the channels) is detected, and the protocol halts without generating any secure keys.
\end{enumerate}

For simplicity, we omit the processes of information reconciliation and privacy amplification. Now we describe the basic BB84 protocol [Steps (1)--(7)] in qCCS as follows.
\begin{eqnarray*}
Alice(n)&\define& \sum_{\tilde{B}_a,
\tilde{K}_a\in \{0,1\}^n}\frac 1{2^{2n}} Set_{\tilde{K}_a}[\tilde{q}].
H_{\tilde{B}_a}[\tilde{q}].\qc{A2B}!\tilde{q}.\Wait_A(\tilde{B}_a,
\tilde{K}_a)
\\
\Wait_A(\tilde{B}_a, \tilde{K}_a)&\define& b2a?\tilde{B}_b.a2b!\tilde{B}_a.
key_a!cmp(\tilde{K}_a, \tilde{B}_a, \tilde{B}_b).\nil
\\
Bob(n)&\define& \qc{A2B}?\tilde{q}. \sum_{\tilde{B}_b\in \{0,1\}^n}\frac 1{2^{n}}M_{\tilde{B}_b}[\tilde{q};\tilde{K}_b].Set_{\tilde{0}}[\tilde{q}].
b2a!\tilde{B}_b.\Wait_B(\tilde{B}_b, \tilde{K}_b)
\\
\Wait_B(\tilde{B}_b, \tilde{K}_b)&\define& a2b?\tilde{B}_a.key_b!cmp(\tilde{K}_b, \tilde{B}_a, \tilde{B}_b).\nil
\\
\mathit{BB84}(n) &\define& Alice(n)\| Bob(n)
\end{eqnarray*}
where $Set_{\tilde{K}_a}[\tilde{q}]$ sets the $i$th qubit of $\tilde{q}$ to the state $|\tilde{K}_a(i)\>$,
 $H_{\tilde{B}_a}[\tilde{q}]$ applies $H$ or does nothing on the $i$th qubit of $\tilde{q}$ depending on whether the $i$th bit of $\tilde{B}_a$ is $1$ or $0$,
and $M_{\tilde{B}_b}[\tilde{q}; \tilde{K}_b]$ is the quantum measurement on $\tilde{q}$ according to the bases determined by $\tilde{B}_b$, i.e., for each $1\leq i\leq n$, it measures $q_i$ with respect to the basis $\{|0\>, |1\>\}$ (resp. $\{|+\>, |-\>\}$) if $\tilde{B}_b(i)=0$ (resp. 1), and stores the result into $\tilde{K}_b(i)$. The function $cmp$ takes a triple of bit-strings $\tilde{x},\tilde{y},\tilde{z}$ with the same size as inputs, and
returns the substring of $\tilde{x}$ where the corresponding bits of $\tilde{y}$ and  $\tilde{z}$ match. When $\tilde{y}$ and  $\tilde{z}$ match nowhere, we let $cmp(\tilde{x}, \tilde{y}, \tilde{z})=\epsilon$, the empty string. We add the operation $Set_{\tilde{0}}[\tilde{q}]$ in $Bob(n)$ for technical reasons: it makes the ideal specifications defined below simple.

To show the correctness of basic {BB84} protocol, we first put $\mathit{BB84}(n)$ in a \emph{test environment} defined as follows
\begin{eqnarray*}
Test &\define& key_a?k_a.key_b?k_b.\iif\ k_a=k_b\ \then\ key!k_a.\nil\ \eelse\ \mathit{fail}!0.\nil\\
\mathit{BB84_{test}}(n) &\define& (\mathit{BB84}(n) \| Test)\backslash \{a2b, b2a, \qc{A2B}, key_a, key_b\}
\end{eqnarray*}
For the ideal \emph{specification} of $\mathit{BB84_{test}}(n)$, we would like it to satisfy the following three conditions: (1) it is correct, in the sense that it will never perform $\mathit{fail}!0$; (2) the generated key $\tilde{x}$ with $|\tilde{x}|=i$ is uniformly distributed for each $i\leq n$. That is, for any $\tilde{x}$ with $|\tilde{x}|=i$, $\mathrm{Pr}(\tilde{x} \mbox{ is the key obtained} \mid\mbox{key-length $= i$}) =  1/2^{i}$; (3) The length of the obtained key
follows the unbiased binomial distribution. That is, for each $i\leq n$, $\mathrm{Pr}(\mbox{key-length $= i$}) = \binom{n}{i}/2^{n}$. Thus we can let
\begin{eqnarray*}
\mathit{BB84_{spec}}(n) &\define &\sum_{i=0}^n \sum_{\tilde{x}\in \{0,1\}^i}\frac {\binom{n}{i}}{2^{n+i}}  Set_{\tilde{0}}[\tilde{q}].key!\tilde{x}.\nil.
\end{eqnarray*}
It is tedious but routine to check that $\mathit{BB84_{test}}(n)\approx \mathit{BB84}_{spec}(n)$ for any $n$. 

 Now we proceed to describe the protocol that detects potential eavesdroppers [Steps (1)--(10)]. Let
 \begin{eqnarray*}
 Alice'(n)&\define & (Alice(n)\|  key_a?\tilde{K}_a'.\sum_{\tilde{x}\subseteq \{1,\dots, m\}}^{|\tilde{x}| = k} \frac 1{\binom{m}{k}}a2b!\tilde{x}.a2b!SubStr(\tilde{K}_a', \tilde{x}).b2a?\tilde{K}_b''.\\
 &&(\iif\ SubStr(\tilde{K}_a', \tilde{x})=\tilde{K}_b''\
  \then\ key'_a!RemStr(\tilde{K}_a', \tilde{x}).\nil))\backslash\{key_a\}
  \\
 Bob'(n)&\define& (Bob(n)\| key_b?\tilde{K}_b'.a2b?\tilde{x}.a2b?\tilde{K}_a''.
 b2a!SubStr(\tilde{K}_b', \tilde{x}).\\
 &&(\iif\ SubStr(\tilde{K}_b', \tilde{x})=\tilde{K}_a''\
  \then\ key'_b!RemStr(\tilde{K}_b', \tilde{x}).\nil))\backslash\{key_b\}\\
\mathit{BB84}'(n) &\define& Alice'(n) \| Bob'(n)  
  \end{eqnarray*}
 where $m=|\tilde{K}_a'|$ and $k=\lceil m/2 \rceil$, the function $SubStr(\tilde{K}_ a', \tilde{x})$ returns the substring of $\tilde{K}_a'$
 at the indexes specified by $\tilde{x}$, and $RemStr(\tilde{K}_a', \tilde{x})$ returns the remaining substring of $\tilde{K}_a'$ by deleting $SubStr(\tilde{K}_a', \tilde{x})$. 
 
 To get a taste of the security of BB84 protocol, 
 we consider a special case where Eve's strategy is to simply measure the qubits sent by Alice, according to randomly guessed  bases, to get the keys and resend these qubits to Bob. 
 That is, we define
  \begin{align*}
 Eve(n) &\define \qc{A2E}? \tilde{q}.\sum_{\tilde{B}_e\in \{0,1\}^n}\frac 1{2^{n}}M_{\tilde{B}_e}[\tilde{q};\tilde{K}_e].key_e'!\tilde{K}_e.\qc{E2B}! \tilde{q}.\nil
 \end{align*}
Again, we put $\mathit{BB84}'(n)$ in a test environment, but now the environment includes the presence of Eve:
\begin{eqnarray*}
Test' &\define&  key'_a?\tilde{x}.key'_b?\tilde{y}.key'_e?\tilde{z}.( \iif\ \tilde{x}\neq \tilde{y}\  \then\ \mathit{fail}!0.\nil\\
&& \hspace{10em} \eelse\ (\iif\ \tilde{x}= \tilde{z}\ \then\ \mathit{hacked}!0.\nil))
\\
\mathit{BB84'_{test}}(n) &\define& (Alice'(n)[f_a] \| Bob'(n)[f_b] \| Eve(n)\| Test')\backslash L
\end{eqnarray*}
where 
$L = \{a2b, b2a, \qc{A2E}, \qc{E2B}, key'_a, key'_b, key'_e\}$, $f_a(\qc{A2B})=\qc{A2E}$, and $f_b(\qc{A2B})=\qc{E2B}$.

Now, to show the \emph{security} of BB84,\footnote{Here we adopt a weak notion of security: by secure we mean the eavesdropper ends up with a false key string. A stronger and more practical notion of security should take into account the mutual information between the keys held by the legitimate parties and the eavesdropper. We leave the analysis of BB84 with respect to this notion of security for future work.} it suffices to prove the following property:
\begin{equation}\label{eq:security}
\mathit{BB84'_{test}}(n) \abisa{c^n} Set_{\tilde{0}}[\tilde{q}].\nil
\end{equation}
where $c=1/2+\sqrt{3}/4<1$. Thus $d_b(\mathit{BB84'_{test}}(n), Set_{\tilde{0}}[\tilde{q}].\nil) \leq c^n$.
That is, the testing system is just like a protocol which only sets the quantum qubits $\tilde{q}$ to $|\tilde{0}\>\<\tilde{0}|$.
As the process $Set_{\tilde{0}}[\tilde{q}].\nil$ never performs $\mathit{fail}!0$ or $\mathit{hacked}!0$, this indicates that the \emph{insecurity degree} of BB84 is at most $c^n$, which decreases exponentially to 0 when $n$ tends to infinity.

To show Eq.(\ref{eq:security}), take arbitrarily $\rho \in \d(\h)$, and let $\c=\<\mathit{BB84'_{test}}(n), \rho\>$ and $\d=\<Set_{\tilde{0}}[\tilde{q}].\nil, \rho\>$.
Basically, we only need to compute the total probability of $\c$ eventually
performing $\mathit{fail}!0$ or $\mathit{hacked}!0$. The reason is, they are the only visible actions of $\c$ ($\d$ does not  perform any visible action at all), and also the only actions which contribute to possible transition inconsistency of distributions obtained from $\c$. If the total probability of their appearance is upper bounded by $c^n$, then $\c$ and $\d$ are $c^n$-bisimilar.

For each qubit sent by Alice, Eve chooses the wrong basis with probability $1/2$, and in this case if Bob measures this qubit according to the correct basis he will get an incorrect result with probability $1/2$. Thus for each qubit that Bob guesses the correct basis, the probability that Alice and Bob get different key bits is $1/4$.
Furthermore, for each $i$-length raw key generated by the basic BB84, Alice and Bob will compare $i/2$ key bits during the eavesdropper-detection phase. The probability that they fail to detect the eavesdropper is then $(3/4)^{i/2}$. 
Note that only when the eavesdropper is not detected, the protocol proceeds. Hence the probability of observing $\mathit{fail}!0$ or $\mathit{hacked}!0$ is upper bounded by 
$$\sum_{i=0}^n \sum_{\tilde{x}\in \{0,1\}^i}\frac {\binom{n}{i}}{2^{n+i}}  (3/4)^{i/2}=\frac 1{2^{n}} \sum_{i=0}^n \binom{n}{i} (3/4)^{i/2}=c^n.$$

\section{Conclusion and Future work}

In this paper, we have proposed a novel notion of distribution-based bisimulation for quantum processes in qCCS. In contrast with previous bisimulations introduced in the literature, our definition is reasonably weaker in that it equates some intuitively bisimilar processes which are not bisimilar according to the previous definitions, thus is more useful in applications. We further defined a bisimulation distance to characterise the extent to which two processes are bisimilar. As an application, we applied the notions of distribution-based bisimulation and bisimulation distance to show that the quantum key distribution protocol BB84 is sound and secure against the intercept-resend attacker. To the best of our knowledge, this is the first time in the literature that the (asymptotic) security of BB84 has been analysed in the framework of a quantum process algebra.

There are still many questions remaining for further study. 
Firstly, as pointed out in Section 6, the notion of security we adopted for the analysis of BB84 is a rather weak one. In quantum information field, people normally use the mutual information between the states held by legitimate parties and the eavesdropper to quantify the leakage of secure information. To perform a security  analysis of BB84 in terms of this stronger notion of security and against more complex model of attack beyond the intercept-resend one studied in the current paper is one of the future directions we are pursuing.

Secondly, bisimilarity checking is usually a very tedious and routine task which can barely be done by hand.
This issue becomes more serious when the number of parties involved and the round of communications increase. To deal with this problem, making the process algebra approach more applicable for the analysis of general quantum cryptographic protocols, we are going to develop a software tool for automated bisimilarity checking. In the theoretical aspect, we will explore the possibility of extending symbolic bisimulation proposed in~\cite{FDY14} to distribution-based case, to decrease the computational complexity of determining bisimilarity.

Finally, as shown in~\cite{Eisentraut:2012wi}, distribution-based bisimulation is not a congruence in general, unless restricted to distributed schedulers.
However, as argued by the authors of~\cite{Eisentraut:2012wi}, non-distributed schedulers, which are responsible for the incongruence, are actually very unrealistic and do not appear in real-world applications. To show that our distribution-based bisimulation is a congruence for qCCS processes under distributed schedulers and to study the implication of  distributed schedulers for quantum cryptographic protocols are also topics worthy of further consideration.

\section*{Acknowledgement}
This work was partially supported by Australian Research Council (Grant No. DP130102764). Y. F. is also supported by  the National Natural Science Foundation of China (Grant Nos. 61428208 and 61472412) and the CAS/SAFEA International Partnership Program for Creative Research Team.

\bibliographystyle{abbrv}

\bibliography{ref}

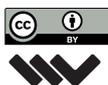
\end{document}